# *Euclid* PREPARATION: XIV. THE COMPLETE CALIBRATION OF THE COLOR-REDSHIFT RELATION (C3R2) SURVEY: DATA RELEASE 3

EUCLID COLLABORATION: S.A. STANFORD[1], D. MASTERS[2], B. DARVISH[3], D. STERN[4], J.G. COHEN[3], P. CAPAK[3], N. HERNITSCHEK[3], I. DAVIDZON[5], J. RHODES[4], D.B. SANDERS[6], B. MOBASHER[7], F.J. CASTANDER[8,9], S. PALTANI[10], N. AGHANIM[11], A. AMARA[12], N. AURICCHIO[13], A. BALESTRA[14], R. BENDER[15,16], C. BODENDORF[16], D. BONINO[17], E. BRANCHINI[18,19,20], J. BRINCHMANN[21,22], V. CAPOBIANCO[17], C. CARBONE[23], J. CARRETERO[24], F. CASAS[8,9], M. CASTELLANO[20], S. CAVUOTI[25,26,27], A. CIMATTI[28], R. CLEDASSOU[29,30], C.J. CONSELICE[31], L. CORCIONE[17], A. COSTILLE[32], M. CROPPER[33], H. DEGAUDENZI[10], M. DOUSPIS[11], F. DUBATH[10], S. DUSINI[34], P. FOSALBA[8,9], M. FRAILIS[35], E. FRANCESCHI[13], P. FRANZETTI[23], M. FUMANA[23], B. GARILLI[23], C. GIOCOLI[13,36], F. GRUPP[15,16], S.V.H. HAUGAN[37], H. HOEKSTRA[38], W. HOLMES[4], F. HORMUTH[39,40], P. HUDELOT[41], K. JAHNKE[40], A. KIESSLING[4], M. KILBINGER[42], T. KITCHING[33], B. KUBIK[43], M. KÜMMEL[15], M. KUNZ[44], H. KURKI-SUONIO[45], R. LAUREIJS[46], S. LIGORI[17], P.B. LILJE[37], I. LLORO[47], E. MAIORANO[13], O. MARGGRAF[48], K. MARKOVIC[4], R. MASSEY[49], M. MENEGHETTI[13,13,36], G. MEYLAN[50], L. MOSCARDINI[13,28,51], S.M. NIEMI[46], C. PADILLA[24], F. PASIAN[35], K. PEDERSEN[52], V. PETTORINO[42], S. PIRES[42], M. PONCET[30], L. POPA[53], L. POZZETTI[13], F. RAISON[16], M. RONCARELLI[13,28], E. ROSSETTI[28], R. SAGLIA[15,16], R. SCARAMELLA[20,54], P. SCHNEIDER[48], A. SECROUN[55], G. SEIDEL[40], S. SERRANO[8,9], C. SIRIGNANO[34,56], G. SIRRI[36], A.N. TAYLOR[57], H.I. TEPLITZ[2], I. TERENO[58,59], R. TOLEDO-MOREO[60], E.A. VALENTIJN[61], L. VALENZIANO[13,36], G.A. VERDES KLEIJN[61], Y. WANG[2], G. ZAMORANI[13], J. ZOUBIAN[55], M. BRESCIA[27], G. CONGEDO[57], L. CONVERSI[62,63], Y. COPIN[43], S. KERMICHE[55], E. MEDINACELI[64], S. MEI[65], M. MORESCO[13,28], B. MORIN[42], E. MUNARI[35], G. POLENTA[66], F. SUREAU[42], P. TALLADA CRESPÍ[67], T. VASSALLO[15], A. ZACCHEI[35], S. ANDREON[68], H. AUSSEL[42], C. BACCIGALUPI[35,69,70,71], A. BALAGUERA-ANTOLÍNEZ[72,73], M. BALDI[13,28,36], S. BARDELLI[13], A. BIVIANO[35,69], E. BORSATO[56], E. BOZZO[10], C. BURIGANA[51,74,75], R. CABANAC[76], S. CAMERA[77,78], A. CAPPI[13,79], C.S. CARVALHO[58], S. CASAS[42], G. CASTIGNANI[50], C. COLODRO-CONDE[73], J. COUPON[10], H.M. COURTOIS[80], J.-G. CUBY[32], A. DA SILVA[59,81], S. DE LA TORRE[32], D. DI FERDINANDO[51], C.A.J. DUNCAN[82], X. DUPAC[63], A.M. FABRICIUS[15,16], M. FARINA[83], S. FARRENS[42], P.G. FERREIRA[82], F. FINELLI[13,51], P. FLOSE-REIMBERG[41], S. FOTOPOULOU[84], S. GALEOTTA[35], K. GANGA[85], W. GILLARD[55], G. GOZALIASL[86], J. GRACIÁ-CARPIO[16], E. KEIHANEN[86], C.C. KIRKPATRICK[45], V. LINDHOLM[86,87], G. MAINETTI[88], D. MAINO[23,89,90], N. MARTINET[32], F. MARULLI[13,36,91], M. MATURI[92,93], S. MAUROGORDATO[79], R. B. METCALF[28,94], R. NAKAJIMA[48], C. NEISSNER[24], J.W. NIGHTINGALE[95], A.A. NUCITA[96,97], L. PATRIZII[36], D. POTTER[98], A. RENZI[34,56], G. RICCIO[27], E. ROMELLI[35], A.G. SÁNCHEZ[16], D. SAPONE[99], M. SCHIRMER[40], M. SCHULTHEIS[79], V. SCOTTEZ[41], L. STANCO[34], M. TENTI[36], R. TEYSSIER[98], F. TORRADEFLOT[24,67], J. VALIVIITA[87,100], M. VIEL[35,69,70,71], L. WHITTAKER[31,101], E. ZUCCA[13]

[1] Department of Physics and Astronomy, University of California, Davis, CA 95616, USA
[2] Infrared Processing and Analysis Center, California Institute of Technology, Pasadena, CA 91125, USA
[3] California institute of Technology, 1200 E California Blvd, Pasadena, CA 91125, USA
[4] Jet Propulsion Laboratory, California Institute of Technology, 4800 Oak Grove Drive, Pasadena, CA, 91109, USA
[5] Cosmic Dawn Center (DAWN), Niels Bohr Institute, University of Copenhagen, Vibenshuset, Lyngbyvej 2, DK-2100 Copenhagen, Denmark
[6] Institute for Astronomy, University of Hawaii, 2680 Woodlawn Drive, Honolulu, HI 96822, USA
[7] Physics and Astronomy Department, University of California, 900 University Ave., Riverside, CA 92521, USA
[8] Institute of Space Sciences (ICE, CSIC), Campus UAB, Carrer de Can Magrans, s/n, 08193 Barcelona, Spain
[9] Institut d'Estudis Espacials de Catalunya (IEEC), Carrer Gran Capità 2-4, 08034 Barcelona, Spain
[10] Department of Astronomy, University of Geneva, ch. dÉcogia 16, CH-1290 Versoix, Switzerland
[11] Université Paris-Saclay, CNRS, Institut d'astrophysique spatiale, 91405, Orsay, France
[12] Institute of Cosmology and Gravitation, University of Portsmouth, Portsmouth PO1 3FX, UK
[13] INAF-Osservatorio di Astrofisica e Scienza dello Spazio di Bologna, Via Piero Gobetti 93/3, I-40129 Bologna, Italy
[14] INAF-Osservatorio Astronomico di Padova, Via dell'Osservatorio 5, I-35122 Padova, Italy
[15] Universitäts-Sternwarte München, Fakultät für Physik, Ludwig-Maximilians-Universität München, Scheinerstrasse 1, 81679 München, Germany
[16] Max Planck Institute for Extraterrestrial Physics, Giessenbachstr. 1, D-85748 Garching, Germany
[17] INAF-Osservatorio Astrofisico di Torino, Via Osservatorio 20, I-10025 Pino Torinese (TO), Italy
[18] INFN-Sezione di Roma Tre, Via della Vasca Navale 84, I-00146, Roma, Italy
[19] Department of Mathematics and Physics, Roma Tre University, Via della Vasca Navale 84, I-00146 Rome, Italy
[20] INAF-Osservatorio Astronomico di Roma, Via Frascati 33, I-00078 Monteporzio Catone, Italy
[21] Centro de Astrofísica da Universidade do Porto, Rua das Estrelas, 4150-762 Porto, Portugal
[22] Instituto de Astrofísica e Ciências do Espaço, Universidade do Porto, CAUP, Rua das Estrelas, PT4150-762 Porto, Portugal
[23] INAF-IASF Milano, Via Alfonso Corti 12, I-20133 Milano, Italy
[24] Institut de Física d'Altes Energies (IFAE), The Barcelona Institute of Science and Technology, Campus UAB, 08193 Bellaterra (Barcelona), Spain
[25] Department of Physics "E. Pancini", University Federico II, Via Cinthia 6, I-80126, Napoli, Italy
[26] INFN section of Naples, Via Cinthia 6, I-80126, Napoli, Italy
[27] INAF-Osservatorio Astronomico di Capodimonte, Via Moiariello 16, I-80131 Napoli, Italy
[28] Dipartimento di Fisica e Astronomia, Università di Bologna, Via Gobetti 93/2, I-40129 Bologna, Italy
[29] Institut national de physique nucléaire et de physique des particules, 3 rue Michel-Ange, 75794 Paris Cédex 16, France
[30] Centre National d'Etudes Spatiales, Toulouse, France
[31] Jodrell Bank Centre for Astrophysics, School of Physics and Astronomy, University of Manchester, Oxford Road, Manchester M13 9PL, UK
[32] Aix-Marseille Univ, CNRS, CNES, LAM, Marseille, France
[33] Mullard Space Science Laboratory, University College London, Holmbury St Mary, Dorking, Surrey RH5 6NT, UK
[34] INFN-Padova, Via Marzolo 8, I-35131 Padova, Italy

[35] INAF-Osservatorio Astronomico di Trieste, Via G. B. Tiepolo 11, I-34131 Trieste, Italy

[36] INFN-Sezione di Bologna, Viale Berti Pichat 6/2, I-40127 Bologna, Italy

[37] Institute of Theoretical Astrophysics, University of Oslo, P.O. Box 1029 Blindern, N-0315 Oslo, Norway

[38] Leiden Observatory, Leiden University, Niels Bohrweg 2, 2333 CA Leiden, The Netherlands

[39] von Hoerner & Sulger GmbH, SchloßPlatz 8, D-68723 Schwetzingen, Germany

[40] Max-Planck-Institut für Astronomie, Königstuhl 17, D-69117 Heidelberg, Germany

[41] Institut d'Astrophysique de Paris, 98bis Boulevard Arago, F-75014, Paris, France

[42] AIM, CEA, CNRS, Université Paris-Saclay, Université de Paris, F-91191 Gif-sur-Yvette, France

[43] Univ Lyon, Univ Claude Bernard Lyon 1, CNRS/IN2P3, IP2I Lyon, UMR 5822, F-69622, Villeurbanne, France

[44] Université de Genève, Département de Physique Théorique and Centre for Astroparticle Physics, 24 quai Ernest-Ansermet, CH-1211 Genève 4, Switzerland

[45] Department of Physics and Helsinki Institute of Physics, Gustaf Hällströmin katu 2, 00014 University of Helsinki, Finland

[46] European Space Agency/ESTEC, Keplerlaan 1, 2201 AZ Noordwijk, The Netherlands

[47] NOVA optical infrared instrumentation group at ASTRON, Oude Hoogeveensedijk 4, 7991PD, Dwingeloo, The Netherlands

[48] Argelander-Institut für Astronomie, Universität Bonn, Auf dem Hügel 71, 53121 Bonn, Germany

[49] Centre for Extragalactic Astronomy, Department of Physics, Durham University, South Road, Durham, DH1 3LE, UK

[50] Observatoire de Sauverny, Ecole Polytechnique Fédérale de Lau- sanne, CH-1290 Versoix, Switzerland

[51] INFN-Bologna, Via Irnerio 46, I-40126 Bologna, Italy

[52] Department of Physics and Astronomy, University of Aarhus, Ny Munkegade 120, DK–8000 Aarhus C, Denmark

[53] Institute of Space Science, Bucharest, Ro-077125, Romania

[54] INFN-Sezione di Roma, Piazzale Aldo Moro, 2 - c/o Dipartimento di Fisica, Edificio G. Marconi, I-00185 Roma, Italy

[55] Aix-Marseille Univ, CNRS/IN2P3, CPPM, Marseille, France

[56] Dipartimento di Fisica e Astronomia "G.Galilei", Università di Padova, Via Marzolo 8, I-35131 Padova, Italy

[57] Institute for Astronomy, University of Edinburgh, Royal Observatory, Blackford Hill, Edinburgh EH9 3HJ, UK

[58] Instituto de Astrofísica e Ciências do Espaço, Faculdade de Ciências, Universidade de Lisboa, Tapada da Ajuda, PT-1349-018 Lisboa, Portugal

[59] Departamento de Física, Faculdade de Ciências, Universidade de Lisboa, Edifício C8, Campo Grande, PT1749-016 Lisboa, Portugal

[60] Universidad Politécnica de Cartagena, Departamento de Electrónica y Tecnología de Computadoras, 30202 Cartagena, Spain

[61] Kapteyn Astronomical Institute, University of Groningen, PO Box 800, 9700 AV Groningen, The Netherlands

[62] European Space Agency/ESRIN, Largo Galileo Galilei 1, 00044 Frascati, Roma, Italy

[63] ESAC/ESA, Camino Bajo del Castillo, s/n., Urb. Villafranca del Castillo, 28692 Villanueva de la Cañada, Madrid, Spain

[64] Istituto Nazionale di Astrofisica (INAF) - Osservatorio di Astrofisica e Scienza dello Spazio (OAS), Via Gobetti 93/3, I-40127 Bologna, Italy

[65] APC, AstroParticule et Cosmologie, Université Paris Diderot, CNRS/IN2P3, CEA/Irfu, Observatoire de Paris, Sorbonne Paris Cité, 10 rue Alice Domon et Léonie Duquet, 75205, Paris Cedex 13, France

[66] Space Science Data Center, Italian Space Agency, via del Politecnico snc, 00133 Roma, Italy

[67] Centro de Investigaciones Energéticas, Medioambientales y Tecnológicas (CIEMAT), Avenida Complutense 40, 28040 Madrid, Spain

[68] INAF-Osservatorio Astronomico di Brera, Via Brera 28, I-20122 Milano, Italy

[69] IFPU, Institute for Fundamental Physics of the Universe, via Beirut 2, 34151 Trieste, Italy

[70] SISSA, International School for Advanced Studies, Via Bonomea 265, I-34136 Trieste TS, Italy

[71] INFN, Sezione di Trieste, Via Valerio 2, I-34127 Trieste TS, Italy

[72] Universidad de la Laguna, E-38206, San Cristóbal de La Laguna, Tenerife, Spain

[73] Instituto de Astrofísica de Canarias. Calle Vía Làctea s/n, 38204, San Cristóbal de la Laguna, Tenerife, Spain

[74] Dipartimento di Fisica e Scienze della Terra, Università degli Studi di Ferrara, Via Giuseppe Saragat 1, I-44122 Ferrara, Italy

[75] INAF, Istituto di Radioastronomia, Via Piero Gobetti 101, I-40129 Bologna, Italy

[76] Institut de Recherche en Astrophysique et Planétologie (IRAP), Université de Toulouse, CNRS, UPS, CNES, 14 Av. Edouard Belin, F-31400 Toulouse, France

[77] INFN-Sezione di Torino, Via P. Giuria 1, I-10125 Torino, Italy

[78] Dipartimento di Fisica, Università degli Studi di Torino, Via P. Giuria 1, I-10125 Torino, Italy

[79] Université Côte d'Azur, Observatoire de la Côte d'Azur, CNRS, Laboratoire Lagrange, Bd de l'Observatoire, CS 34229, 06304 Nice cedex 4, France

[80] University of Lyon, UCB Lyon 1, CNRS/IN2P3, IUF, IP2I Lyon, France

[81] Instituto de Astrofísica e Ciências do Espaço, Faculdade de Ciências, Universidade de Lisboa, Campo Grande, PT-1749-016 Lisboa, Portugal

[82] Department of Physics, Oxford University, Keble Road, Oxford OX1 3RH, UK

[83] INAF-Istituto di Astrofisica e Planetologia Spaziali, via del Fosso del Cavaliere, 100, I-00100 Roma, Italy

[84] School of Physics, HH Wills Physics Laboratory, University of Bristol, Tyndall Avenue, Bristol, BS8 1TL, UK

[85] Université de Paris, CNRS, Astroparticule et Cosmologie, F-75006 Paris, France

[86] Department of Physics, P.O. Box 64, 00014 University of Helsinki, Finland

[87] Helsinki Institute of Physics, Gustaf Hällströmin katu 2, University of Helsinki, Helsinki, Finland

[88] Centre de Calcul de l'IN2P3, 21 avenue Pierre de Coubertin F-69627 Villeurbanne Cedex, France

[89] Dipartimento di Fisica "Aldo Pontremoli", Università degli Studi di Milano, Via Celoria 16, I-20133 Milano, Italy

[90] INFN-Sezione di Milano, Via Celoria 16, I-20133 Milano, Italy

[91] Dipartimento di Fisica e Astronomia "Augusto Righi" - Alma Mater Studiorum Università di Bologna, via Piero Gobetti 93/2, I-40129 Bologna, Italy

[92] Institut für Theoretische Physik, University of Heidelberg, Philosophenweg 16, 69120 Heidelberg, Germany

[93] Zentrum für Astronomie, Universität Heidelberg, Philosophenweg 12, D- 69120 Heidelberg, Germany

[94] INAF-IASF Bologna, Via Piero Gobetti 101, I-40129 Bologna, Italy

[95] ICC&CEA, Department of Physics, Durham University, South Road, DH1 3LE, UK

[96] INFN, Sezione di Lecce, Via per Arnesano, CP-193, I-73100, Lecce, Italy

[97] Department of Mathematics and Physics E. De Giorgi, University of Salento, Via per Arnesano, CP-I93, I-73100, Lecce, Italy

[98] Institute for Computational Science, University of Zurich, Winterthurerstrasse 190, 8057 Zurich, Switzerland

[99] Departamento de Física, FCFM, Universidad de Chile, Blanco Encalada 2008, Santiago, Chile

[100] Department of Physics, P.O.Box 35 (YFL), 40014 University of Jyväskylä, Finland

[101] Department of Physics and Astronomy, University College London, Gower Street, London WC1E 6BT, UK

## ABSTRACT

The Complete Calibration of the Color-Redshift Relation (C3R2) survey is obtaining spectroscopic redshifts in order to map the relation between galaxy color and redshift to a depth of $i \sim 24.5$ (AB). The primary goal is to enable sufficiently accurate photometric redshifts for Stage IV dark energy projects, particularly *Euclid* and the *Roman Space Telescope*, which are designed to constrain cosmological parameters through weak lensing. We present 676 new high-confidence spectroscopic redshifts obtained by the C3R2 survey in the 2017B–2019B semesters using the DEIMOS, LRIS, and MOSFIRE multi-object spectrographs on the Keck telescopes. Combined with the 4454 redshifts previously published by this project, the C3R2 survey has now obtained and published 5130 high-quality galaxy spectra and redshifts. If we restrict consideration to only the $0.2 < z_\mathrm{p} < 2.6$ range of interest for the *Euclid* cosmological goals, then with the current data release C3R2 has increased the spectroscopic redshift coverage of the *Euclid* color space from 51% (as reported by Masters et al. 2015) to the current 91%. Once completed and combined with extensive data collected by other spectroscopic surveys, C3R2 should provide the spectroscopic calibration set needed to enable photometric redshifts to meet the cosmology requirements for *Euclid*, and make significant headway toward solving the problem for *Roman*.

*Keywords:* galaxies — surveys: spectroscopic

## 1. INTRODUCTION

The Stage IV cosmology projects, such as the Vera C. Rubin Observatory Legacy Survey of Space and Time (hereafter LSST), *Euclid*, and the *Nancy Grace Roman Space Telescope* (hereafter *Roman*), will use deep imaging in multiple optical and near-infrared filters to calculate photometric redshifts ($z_\mathrm{p}$) of large numbers of galaxies for weak lensing cosmology. While high-quality photometric redshifts are crucial for many other investigations, the photometric redshift requirements for cosmology are the most exacting. Weak lensing cosmology requires *unbiased* redshift estimates (Ma et al. 2006; Huterer et al. 2006), that are very challenging for photometric redshift estimation techniques. The requirements for *Euclid* and *Roman* are that the mean redshift ⟨z⟩ of galaxies in ~10 tomographic redshift bins for cosmic shear measurements must be known to better than 0.2% (that is, $\Delta\langle z \rangle \leq 0.002(1 + \langle z \rangle)$ for each bin) in the $0.2 < z < 2.6$ range. Numerous tests have shown that this level of accuracy is not currently feasible for photometric redshift algorithms and realistic training samples (Desprez et al. 2020). The bias requirement, in particular, makes spectroscopic calibration samples necessary for the success of these missions (Newman et al. 2013).

**Table 1**. Overview of the deep fields targeted by C3R2 in DR3.

| Field | Right Ascension (J2000) | Declination (J2000) | Area (deg$^2$) | Optical data (*ugriz*) | Near-IR data (*YJHK$_s$*) |
|---|---|---|---|---|---|
| VVDS-2h | 02$^h$ 26$^m$ | −04° 30′ | 1.0 | CFHTLS | VISTA |
| COSMOS | 10$^h$ 00$^m$ | +02° 12′ | 2.0 | CFHTLS | VISTA |
| EGS | 14$^h$ 19$^m$ | +52° 41′ | 1.0 | CFHTLS | CFHTLS-WIRDS[a] |

[a] *Y*-band obtained from CFHT-WIRCAM observations separate from the WIRDS survey.

Stage III cosmology surveys currently underway also face the photometric redshift estimation challenge, and constitute an important testbed for the photometric redshift techniques to be employed in the Stage IV projects. These include the Kilo-Degree Survey (KiDS; de Jong et al. 2015, 2017; Hildebrandt et al. 2017), the Dark Energy Survey (DES; DES Collaboration et al. 2018; Troxel et al. 2018), and the Hyper-Suprime Cam (HSC) Survey (Aihara et al. 2018). A variety of techniques have been employed in these surveys to constrain redshift distributions, $N(z)$, of galaxies in shear bins – such as clustering redshifts which use the spatial distribu-tion of overlapping spectroscopic redshift samples to infer the $N(z)$ distribution of the photometric sample (e.g., Newman 2008; Ménard et al. 2013; McQuinn & White 2013; Schmidt et al. 2013; Morrison et al. 2017), template fitting (e.g., Benítez 2000; Brammer et al. 2008; Ilbert et al. 2009), and machine learning based on training samples (e.g., Collister & Lahav 2004; Carrasco Kind & Brunner 2013). Reweighting of the spectroscopic sample to better match the photometric sample, as described in Lima et al. (2008), has also been used. It is generally agreed that a key limitation for the empirical photo-z and re-weighting techniques is the need for a fully representative calibration sample of spectroscopic redshifts ($z_s$) that explores the range of galaxy properties present in the surveys.

The Complete Calibration of the Color-Redshift Relation (C3R2) survey (Masters et al. 2017, hereafter M17 [DR1]) was initiated in response to this need, with the goal of mapping the empirical relation between galaxy redshift and color to the *Euclid* depth of $i \sim 24.5$ (AB). The survey strategy is based on the fact that the observed galaxy colors down to a given survey depth are both *limited* and *measurable*. Moreover, a well-defined, mostly non-degenerate relation between a galaxy's position in multi-color space ($C$) and its redshift ($z$) can be discovered empirically. Elucidating this P($z|C$) relation is the goal of the C3R2 survey.

The C3R2 survey strategy follows the method outlined in (Masters et al. 2015, hereafter M15). M15 illustrated the use of an unsupervised manifold learning algorithm, the *self-organizing map* (SOM; Kohonen 1982), to map the color distribution of galaxies in the high-dimensional color space (*u-g, g-r, ..., J-H*) anticipated for *Euclid* and *Roman* photometric redshift estimation. The SOM is a neural network model widely used to map and identify correlations in high-dimensional data. The algorithm uses unsupervised, competitive learning of "neurons" to project high-dimensional data onto a lower-dimensional grid. The SOM algorithm can be thought of as a type of nonlinear principal component analysis, and is also similar in some respects to the k-means clustering algorithm. In contrast to these and other methods, the SOM preserves the topology of the high-dimensional data in the low-dimension representation. Similar objects are thus grouped together on the self-organized map, and clusters that exist in the high-dimensional data space are reflected in the lower-dimensional representation. This feature makes the maps visually understandable and thus useful for identifying correlations that exist in high-dimensional data. This high-dimensional mapping allows us to directly determine which parts of galaxy color space are well sampled with existing spectroscopy and which are not, thus letting us focus spectroscopic calibration effort on those regions which are the least constrained. We have been focussed on obtaining only the necessary additional spectroscopy needed to build a representative calibration sample, such that direct inference of the P($z|C$) relation can be made sufficiently accurate to meet the cosmology requirements.

In M17 and (Masters et al. 2019, hereafter M19, i.e. DR2) we presented the results of the C3R2 Keck observations taken in the 2016A, 2016B, and 2017A semesters. Here we present results obtained in the 2017B, 2018A, 2018B, and 2019B semesters, comprising 18 Keck nights (some of which were shared, or half-nights), of which ~11 had good observing conditions. The new data come from nights allocated by NASA to PI D. Stern (2 nights) and to PI D. Masters (16 nights). We refer the reader to M15 for background on the calibration approach, and to M17 and M19 for the observation and data reduction details of the C3R2 survey. Spectroscopic redshifts have also been obtained using the VLT as part of the C3R2 effort (Guglielmo et al. 2020).

This paper is structured as follows. In §2 we describe the observations and data reduction. The observations were all conducted in the fields listed in Table 1. In §3 we present our spectroscopic results from 19 nights of Keck time in semesters 2017B–2020A, discuss the performance of the method, and investigate the status of the calibration effort and the issues still to be addressed.

## 2. OBSERVATIONS AND DATA REDUCTIONS

The observations were carried out at the Keck Observatory located on Mauna Kea in Hawaii. Both of the two 10 m telescopes were used, as LRIS and MOSFIRE are located on Keck 1 and DEIMOS on Keck 2. The observing nights and weather conditions are summarized in Table 2. Note that approximately 6 of the 19 nights were lost to bad weather. Table 3 summarizes the observed slit masks.

### 2.1. *Description of Observations*

Observations for DR3 were carried out essentially in the same manner as described in M17 and M19 for the previous data releases. As discussed in more detail in M17 and M19, we estimated the exposure times needed to obtain redshifts for each target with the three main instruments available at Keck, i.e. DEIMOS, LRIS, and MOSFIRE, and assigned targets to the instrument with the lowest expected exposure time. Here we give a brief overview of the instruments and methods used in the observing.

#### 2.1.1. *DEep Imaging Multi-Object Spectrograph (DEIMOS)*

DEIMOS (Faber et al. 2003) is a wide-field optical spectrograph that covers a roughly rectangular field of view measuring $16.5 \times 5$ arcmin$^2$. Custom slit masks typically target ~100 galaxies at the same time, providing spectra from 5000 Å up to 1 $\mu m$ at spectral resolution $R \equiv \lambda/\Delta\lambda \sim 3000$. We used the 600 groove mm$^{-1}$ grating blazed at 7200 Å and the GG400 blocking filter, with dithering performed to improve sky subtraction. We use a minimum slit length of 8$''$ as a balance between maximizing the number of targets on the mask and getting good sky measurements. Data were reduced using a modified version of the DEEP2 pipeline designed to deal with dithered data.

#### 2.1.2. *Multi-Object Spectrograph for Infrared Exploration (MOSFIRE)*

MOSFIRE (McLean et al. 2012) is a near-IR spectrograph which can observe up to 46 objects at the same time over a $6.1 \times 6.1$ arcmin$^2$ field. Data may be obtained in one of the *YJHK* bandpasses at typical resolutions of $R \sim 3000$. We used MOSFIRE in its default configuration. For instrumental details we refer the reader to Steidel et al. (2014). For our *H*-band observations we used integration times of 120 s with ABAB dithering to improve sky subtraction. Reductions were performed with the MOSFIRE Data Reduction Pipeline (DRP) made available by the instrument team[1]. We chose to observe only in *H*-band because the density of high-priority targets for which we could expect to get a secure redshift was notably higher in this band; this is largely due to the limited wavelength range of a single bandpass in MOSFIRE multislit spectroscopy, the $z \sim 1.5$ redshifts of high-priority targets, and the resulting observed-frame wavelengths of prominent emission lines.

---

[1] https://keck-datareductionpipelines.github.io/MosfireDRP/

### 2.1.3. *Low-Resolution Imaging Spectrograph (LRIS)*

LRIS (Oke et al. 1995) is a dual beam optical spectrograph which can simultaneously cover from ~3200 Å to 1 $\mu m$ over a $6 \times 7.8$ arcmin$^2$ field. Spectra with resolution in the range $300 < R < 5000$ may be obtained using custom masks which typically allow one to observe ~25 galaxies at the same time.

We used LRIS with the 400 groove mm$^{-1}$ blue grism blazed at 3400 Å and the 400 groove mm$^{-1}$ red grating blazed at 8500 Å, with the D560 dichroic. Our choice of blue grism gives high sensitivity at bluer wavelengths where spectral features are likely to be found for objects with photometric redshifts of $z \sim 1.5 - 3$, while the red coverage allows for the detection of [O II] $\lambda3727$ for sources out to $z \sim 1.6$. Note that the blue camera was inoperative on the night of UT 2019 January 2. The LRIS spectra were reduced using the IRAF-based `BOGUS` software developed by D. Stern, S. A. Stanford, and A. Bunker, and flux calibrated using observations of standard stars from Massey & Gronwall (1990) observed on the same night using the same instrument configuration.

**Table 2.** List of observing nights.

| UT Date | Code | Instrument | # Masks | Notes |
|---|---|---|---|---|
| 2017 December 11 | N30-D | DEIMOS | 2.3 | light cirrus; $0\rlap{.}''7 - 1\rlap{.}''3$ seeing |
| 2017 December 12 | N31-D | DEIMOS | 2.3 | $0\rlap{.}''8 - 2\rlap{.}''3$ seeing |
| 2017 December 13 | N32-D | DEIMOS | 3.3 | intermittent clouds |
| 2018 January 11 | N33-D | DEIMOS | 2 | thin cirrus to thick clouds; seeing $1\rlap{.}''1 - 2\rlap{.}''1$ |
| 2018 January 12 | N34-D | DEIMOS | 3 | thin cirrus; $1\rlap{.}''3 - 1\rlap{.}''7$ seeing |
| 2018 February 9 | N35-M | MOSFIRE | 4 | $1\rlap{.}''0 - 1\rlap{.}''3$ seeing |
| 2018 February 28 | N36-M | MOSFIRE | 2 | thick clouds 1st half; clear and $0\rlap{.}''8$ seeing 2nd half |
| 2018 April 3 | N37-M | MOSFIRE | $\cdots$ | lost to weather: fog/humidity |
| 2018 April 5 | N38-M | MOSFIRE | $\cdots$ | lost to weather: fog/humidity |
| 2018 April 6 | N39-M | MOSFIRE | $\cdots$ | lost to weather: fog/humidity |
| 2018 December 6 (first half) | N40-L | LRIS | 2 | clear, windy; $0\rlap{.}''8 - 1\rlap{.}''0$ seeing |
| 2018 December 7 | N41-L | LRIS | 4 | clear; $0\rlap{.}''8 - 1\rlap{.}''2$ seeing |
| 2018 December 31 (first half) | N42-L | LRIS | 2 | lost to weather: thick clouds, $2\rlap{.}''0$ seeing |
| 2019 January 1 (first half) | N43-L | LRIS | 3.5 | clear; $1\rlap{.}''0 - 1\rlap{.}''5$ seeing |
| 2019 January 2 | N44-L | LRIS | 5 | clear; $1\rlap{.}''0 - 1\rlap{.}''4$ seeing; **blue camera inoperative** |
| 2019 January 29 | N45-L | LRIS | 1 | high wind; poor seeing; closed early |
| 2019 January 30 (first half) | N46-L | LRIS | 1 | lost to weather: ice/wind/moisture |
| 2019 December 11 | N47-M | MOSFIRE | 4 | clear; $0\rlap{.}''6$ seeing |
| 2019 December 31 | N48-M | MOSFIRE | 4 | clear; $0\rlap{.}''6 - 1\rlap{.}''0$ seeing |
| 2020 January 19 | N49-M | MOSFIRE | 4 | clear; $0\rlap{.}''5$ seeing |
| 2020 October 18 (2nd half) | N50-L | LRIS | 2 | partly cloudy; $1\rlap{.}''0$ seeing |
| 2020 October 19 (2nd half) | N51-L | LRIS | 2 | partly cloudy; $1\rlap{.}''0$ seeing |

### 2.2. *Redshift determination*

We refer the reader to M17 for a detailed description of the redshift determination procedure, as well as the quality flags and failure codes we adopt. To summarize, each observed source was assessed independently by two co-authors to determine the redshift and associated quality flag ($Q = 0 - 4$) defined as follows:

- $Q = 4$: A quality flag of 4 indicates an unambiguous redshift identified with multiple features or the presence of the split [O II] $\lambda3727$ doublet.

- $Q = 3.5$: A quality flag of 3.5 indicates a high-confidence redshift based on a single line, with a remote possibility of an incorrect identification. An example might be a strong, isolated emission line identified as H$\alpha$, where other identifications of the line are highly improbable due to the lack of associated lines or continuum breaks. This flag is typically only adopted for LRIS and MOSFIRE spectra; single line redshifts in DEIMOS spectra are usually the O II doublet which is split by the DEIMOS spectral resolution.

- $Q = 3$: A quality flag of 3 indicates a high-confidence redshift with a low probability of an incorrect identifi-

cation. An example might be the low signal-to-noise ratio detection of an emission line, possibly corrupted by telluric emission or absorption, identified as [O II] $\lambda3727$, but where the data quality is insufficient to clearly resolve the doublet.

- $Q = 2/1$: A quality flag of 2 indicates a reasonable guess, while a quality flag of 1 indicates a highly uncertain guess. Sources with these low-confidence redshifts are not included in the data release.

- $Q = 0$: A quality flag of 0 indicates that no redshift could be identified. As described above, a code indicating the cause of the redshift failure is assigned in place of the redshift.

These flags are meant to be qualitative indicators. Sample spectra associated with each flag are given in Figure 2 of DR1. Any conflicting assessments were reconciled through a joint review of the spectra with the help of a third, independent reviewer. Sources for which we failed to identify a redshift were assigned $Q = 0$ and a failure code to indicate the most likely reason:

- Code = −91: Insufficient S/N;
- Code = −92: Well-detected but no discernible features;
- Code = −93: Problem with the reduction;
- Code = −94: Missing slit (an extreme case of −93).

As in M17, we also investigated all $Q = 4$ sources for which the spectroscopic redshift ($z_s$) was highly discrepant from the expected photometric redshift ($z_p$), where the latter is a SOM-based value calculated by combining individual photometric redshifts for all galaxies in a SOM cell, as determined with the template-fitting *Le Phare* code (Arnouts et al. 1999; Ilbert et al. 2006) to deep 30-band photometry. The results of this analysis are presented in Section 3.3.

**Table 3**. List of observed slitmasks.

| Mask ID / Name | Night(s) | R.A. (J2000) | Dec (J2000) | P.A. (°) | Exposure (s) | # Spectra (targets / Q = 4 / Q >= 3 serendips) |
|---|---|---|---|---|---|---|
| 17B-D094 / VVDS32 | N30-D | $02^h\,27^m\,22\overset{s}{.}01$ | −04° 48′ 02″ | +90.0 | 15×1200 | 82 / 17 / 2 |
| 17B-D095 / COSMOS-1hr1 | N30,N31,N32-D | $09^h\,59^m\,44\overset{s}{.}20$ | +02° 35′ 17″ | +90.0 | 3×1200 | 80 / 29 / 2 |
| 17B-D096 / COSMOS-3hr1 | N30-D | $09^h\,58^m\,25\overset{s}{.}20$ | +01° 43′ 16″ | +90.0 | 11×1200 | 81 / 53 / 7 |
| 17B-D097 / VVDS33 | N31-D | $02^h\,25^m\,40\overset{s}{.}40$ | −04° 03′ 11″ | +90.0 | 15×1200 | 77 / 16 / 3 |
| 17B-D098 / COSMOS-3hr2 | N31-D | $09^h\,59^m\,57\overset{s}{.}50$ | +01° 55′ 45″ | +90.0 | 11×1200 | 88 / 6 / 0 |
| 17B-D099 / VVDS34 | N32-D | $02^h\,25^m\,30\overset{s}{.}10$ | −04° 31′ 58″ | +90.0 | 9×1200 | 79 / 16 / 3 |
| 17B-D100 / VVDS27 | N32-D | $02^h\,25^m\,19\overset{s}{.}50$ | −04° 42′ 41″ | +90.0 | 3×1200 | 80 / 24 / 2 |
| 17B-D101 / COSMOS-3hr3 | N32-D | $10^h\,00^m\,36\overset{s}{.}70$ | +01° 39′ 47″ | +90.0 | 12×1200 | 84 / 34 / 5 |
| 17B-D102 / VVDS35 | N33-D | $02^h\,24^m\,32\overset{s}{.}30$ | −04° 30′ 23″ | +90.0 | 9×1200 | 81 / 31 / 1 |
| 17B-D103 / COSMOS-3hr4 | N33-D | $09^h\,58^m\,44\overset{s}{.}10$ | +01° 51′ 40″ | +90.0 | 4×1200 | 81 / 2ª / 0 |
| 17B-D104 / VVDS36 | N34-D | $02^h\,24^m\,33\overset{s}{.}30$ | −04° 37′ 23″ | +90.0 | 10×1200 | 78 / 25 / 4 |
| 17B-D105 / COSMOS-3hr6 | N34-D | $10^h\,00^m\,28\overset{s}{.}80$ | +02° 42′ 00″ | +90.0 | 9×1200 | 85 / 33 / 0 |
| 17B-D106 / COSMOS-3hr7 | N34-D | $10^h\,01^m\,12\overset{s}{.}00$ | +01° 39′ 00″ | +90.0 | 9×1200 | 89 / 35 / 2 |
| 18A-M107 / VVDS-H-m1 | N35-M | $02^h\,24^m\,26\overset{s}{.}50$ | −04° 05′ 27″ | +55.0 | 30×120 | 23 / 0 / 0 |
| 18A-M108 / COSMOS-H-2hr5 | N35-M | $10^h\,01^m\,28\overset{s}{.}50$ | +01° 52′ 20″ | +45.0 | 60×120 | 15 / 0 / 4 |
| 18A-M109 / COSMOS-H-2hr1 | N35-M | $09^h\,59^m\,15\overset{s}{.}70$ | +02° 43′ 05″ | +5.0 | 60×120 | 18 / 9 / 3 |
| 18A-M110 / COSMOS-H-2hr3 | N35-M | $10^h\,01^m\,11\overset{s}{.}80$ | +02° 19′ 00″ | −10.0 | 60×120 | 19 / 1 / 0 |
| 18A-M111 / EGS_H_1hr1 | N36-M | $14^h\,16^m\,30\overset{s}{.}00$ | +52° 54′ 00″ | +15.0 | 44×120 | 20 / 3 / 0 |
| 18A-M112 / EGS_H_1hr2 | N36-M | $14^h\,18^m\,10\overset{s}{.}00$ | +52° 38′ 30″ | −25.0 | 46×120 | 15 / 0 / 0 |
| 18B-L113 / VVDS_1h1 | N40-L | $02^h\,25^m\,19\overset{s}{.}20$ | −04° 02′ 38″ | 0.0 | 3×1200 | 29 / 2 / 0 |
| 18B-L114 / VVDS_3h1 | N40-L | $02^h\,24^m\,59\overset{s}{.}90$ | −04° 06′ 47″ | 0.0 | 4×1200 | 28 / 2ª / 3 |
| 18B-L115 / VVDS_3h3 | N41-L | $02^h\,25^m\,57\overset{s}{.}12$ | −04° 49′ 19″ | 0.0 | 9×1200 | 33 / 2 / 0 |
| 18B-L116 / VVDS_1h2 | N41-L | $02^h\,24^m\,26\overset{s}{.}32$ | −04° 10′ 19″ | 0.0 | 4(5)×1200B(R) | 32 / 7 / 0 |
| 18B-L117 / VVDS_1h3 | N41-L | $02^h\,24^m\,13\overset{s}{.}20$ | −04° 44′ 13″ | 0.0 | 2×1200 | 29 / 0 / 0 |
| 18B-L118 / COSMOS_3h1 | N41-L | $10^h\,01^m\,28\overset{s}{.}32$ | +02° 21′ 04″ | −29.0 | 9×1200 | 31 / 0 / 0 |
| 18B-L119 / VVDS_1h4 | N43-L | $02^h\,26^m\,43\overset{s}{.}47$ | −04° 43′ 41″ | 0.0 | 1×3600 | 32 / 0 / 0 |
| 18B-L120 / VVDS_1h5 | N43-L | $02^h\,27^m\,19\overset{s}{.}90$ | −04° 51′ 40″ | 0.0 | 1×3600 | 31 / 3 / 1 |
| 18B-L121 / VVDS_1h6 | N43-L | $02^h\,26^m\,14\overset{s}{.}44$ | −04° 39′ 38″ | 0.0 | 1×3600 | 31 / 0 / 1 |
| 18B-L122 / VVDS_3h2 | N43-L | $02^h\,24^m\,39\overset{s}{.}98$ | −04° 20′ 20″ | 0.0 | 1×3600 | 29 / 0ª / 0 |

Table 3 *(continued)*

| Mask ID / Name | Night(s) | R.A. (J2000) | Dec (J2000) | P.A. ($^{o}$) | Exposure (s) | # Spectra (targets / $Q = 4$ / $Q >= 3$ serendips) |
|---|---|---|---|---|---|---|
| 18B-L123 / VVDS_3h5 | N44-L | $02^h 27^m 31\overset{s}{.}33$ | $-04° 43' 57''$ | 0.0 | 8×1260 | 32 / 2[b] / 0 |
| 18B-L124 / VVDS_1h7 | N44-L | $02^h 27^m 29\overset{s}{.}52$ | $-04° 32' 10''$ | +90 | 3×1200 | 26 / 0[b] / 0 |
| 18B-L125 / COSMOS_1h3 | N44-L | $10^h 01^m 36\overset{s}{.}00$ | $+02° 02' 24''$ | 0.0 | 3×1200 | 24 / 1[b] / 0 |
| 18B-L126 / COSMOS_1h4 | N44-L | $10^h 00^m 19\overset{s}{.}20$ | $+02° 42' 00''$ | 0.0 | 3×1200 | 26 / 0[b] / 0 |
| 18B-L127 / COSMOS_3h2 | N44-L | $10^h 00^m 27\overset{s}{.}33$ | $+02° 23' 20''$ | 0.0 | 5×1200 | 29 / 2[b] / 0 |
| 19B-M128 / VVDS-H-3h1 | N47-M | $02^h 27^m 25\overset{s}{.}98$ | $-04° 47' 17''$ | −60.0 | 78×120 | 23 / 6 / 1 |
| 19B-M129 / VVDS-H-2h2 | N47-M | $02^h 27^m 05\overset{s}{.}82$ | $-04° 17' 31''$ | −70.0 | 63×120 | 21 / 2 / 2 |
| 19B-M130 / COSMOS-H-1h1 | N47-M | $10^h 01^m 22\overset{s}{.}26$ | $+01° 51' 45''$ | −65.0 | 31×120 | 17 / 9 / 0 |
| 19B-M131 / COSMOS-H-3h3 | N47-M | $10^h 00^m 33\overset{s}{.}68$ | $+01° 17' 56''$ | +10.0 | 51×120 | 16 / 7 / 4 |
| 19B-M132 / VVDS-H-3h3 | N48-M | $02^h 25^m 15\overset{s}{.}00$ | $-04° 52' 35''$ | +55.0 | 80×120 | 24 / 6 / 3 |
| 19B-M133 / VVDS-H-2h3 | N48-M | $02^h 24^m 13\overset{s}{.}53$ | $-04° 54' 40''$ | +350.0 | 40×120 | 21 / 1 / 0 |
| 19B-M134 / COSMOS-H-2h4 | N48-M | $10^h 01^m 29\overset{s}{.}88$ | $+01° 48' 56''$ | +350.0 | 56×120 | 19 / 4 / 0 |
| 19B-M135 / COSMOS-H-2h2 | N48-M | $09^h 58^m 33\overset{s}{.}90$ | $+02° 11' 07''$ | +350.0 | 55×120 | 17 / 8 / 2 |
| 19B-M136 / VVDS-H-1h3 | N49-M | $02^h 27^m 28\overset{s}{.}53$ | $-04° 51' 52''$ | −10.0 | 20×120 | 22 / 1 / 3 |
| 19B-M137 / COSMOS-H-2h3 | N49-M | $10^h 00^m 16\overset{s}{.}79$ | $+02° 41' 20''$ | +50.0 | 48×120 | 20 / 9 / 0 |
| 19B-M138 / COSMOS-H-3h1 | N49-M | $10^h 00^m 41\overset{s}{.}61$ | $+01° 45' 50''$ | +119.0 | 80×120 | 18 / 6 / 2 |
| 19B-M139 / COSMOS-H-1h3 | N49-M | $09^h 58^m 05\overset{s}{.}91$ | $+02° 27' 12''$ | +20.0 | 28×120 | 16 / 3[a] / 0 |
| 20B-L128 / VVDS_1hr3 | N50-L | $02^h 25^m 23\overset{s}{.}92$ | $-04° 14' 53''$ | −11.5 | 6×1200 | 25 / 6 / 0 |
| 20B-L129 / VVDS_1hr5 | N50-L | $02^h 26^m 11\overset{s}{.}88$ | $-04° 06' 07''$ | +13.0 | 7×1200 | 26 / 6 / 3 |
| 20B-L130 / VVDS_1hr2 | N51-L | $02^h 24^m 38\overset{s}{.}10$ | $-04° 56' 10''$ | +69.0 | 6×1200 | 29 / 5 / 3 |
| 20B-L131 / VVDS_1hr1 | N51-L | $02^h 24^m 37\overset{s}{.}28$ | $-04° 14' 59''$ | −7.0 | 6×1200 | 28 / 2 / 2 |

NOTE—'Night' column refers to observing code in second column of Table 2: night number, followed by letter indicating instrument used (D – DEIMOS, L – LRIS, M – MOSFIRE). R.A. and Dec refer to the mask center. Final column gives total number of slitlets in mask, total number of highest-quality ($Q = 4$) redshifts measured, and the number of serendipitous sources with highest-quality redshifts ($Q = 4$).

[a] Exposure time significantly less than planned.

[b] No blue side data obtained.

## 3. REDSHIFT RESULTS

In this data release we targeted 1959 objects in 50 masks; 18/19/13 were observed with MOSFIRE/LRIS/DEIMOS respectively. We were able to obtain 435 redshifts with $Q = 4$ and 174 redshifts with either $Q = 3$ or $Q = 3.5$ for the targeted objects; these numbers do not include serendipitous sources which are discussed below. Thus the success rate for all high-confidence redshifts ($Q \geq 3$) is 31% of targeted objects. If we restrict the results to the data obtained in good conditions (32 masks) then we obtained high-confidence redshifts for 511 out of 1346 targeted objects for a success rate of 38%. We return to a discussion of the relatively low success rate in Section 3.4.

Combined with the previous DR1 and DR2 results, C3R2 now has obtained 5131 high-confidence spectroscopic redshifts. The updated redshift and magnitude distributions of the entire spectroscopic sample are presented in Figure 1. The newly reported DR3 redshifts are available in a machine-readable table, of which the first few lines are presented in Table 4 in the Appendix.

### 3.1. *Redshift Performance of the SOM Technique*

Figures 2 through 4 show comparisons of highest quality ($Q = 4$) spectroscopic redshifts with the predicted SOM photometric redshifts (i.e. prior to obtaining the new spectroscopic redshifts) for the new DR3 redshifts (Figure 2), for the combination of all three C3R2 data releases (Figure 3), and for a combination of all our C3R2 redshifts with high-quality literature redshifts for only the COSMOS field (Figure 4). For the latter plot, we use the median of the *Le Phare* photometric redshifts (Laigle et al. 2016) of the galaxies in each cell as the SOM photometric redshift for all galaxies in a cell.

For the case of all our high-quality C3R2 redshifts , as seen in Figure 3, we calculate the scatter, defined as the normalized median absolute deviation

$$\sigma_{NMAD} = 1.48 \times median \left( \frac{|(z_p - z_s)|}{1 + z_s} \right)$$

to be 0.023, the true outlier fraction (defined as the sources with $|z_p - z_s|/(1 + z_s) > 0.15$) to be 0.041, and the bias

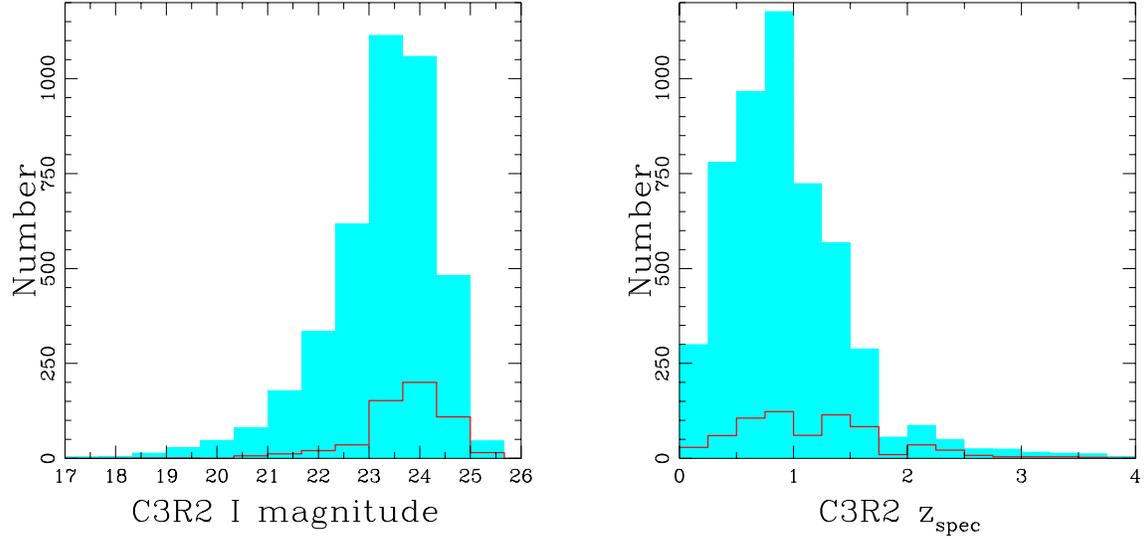

**Figure 1**. Magnitude and redshift histograms of the objects with Q ≥ 3 redshifts. The shaded cyan histograms represent objects in the first three C3R2 data releases, and the red histograms represent only objects in DR3.

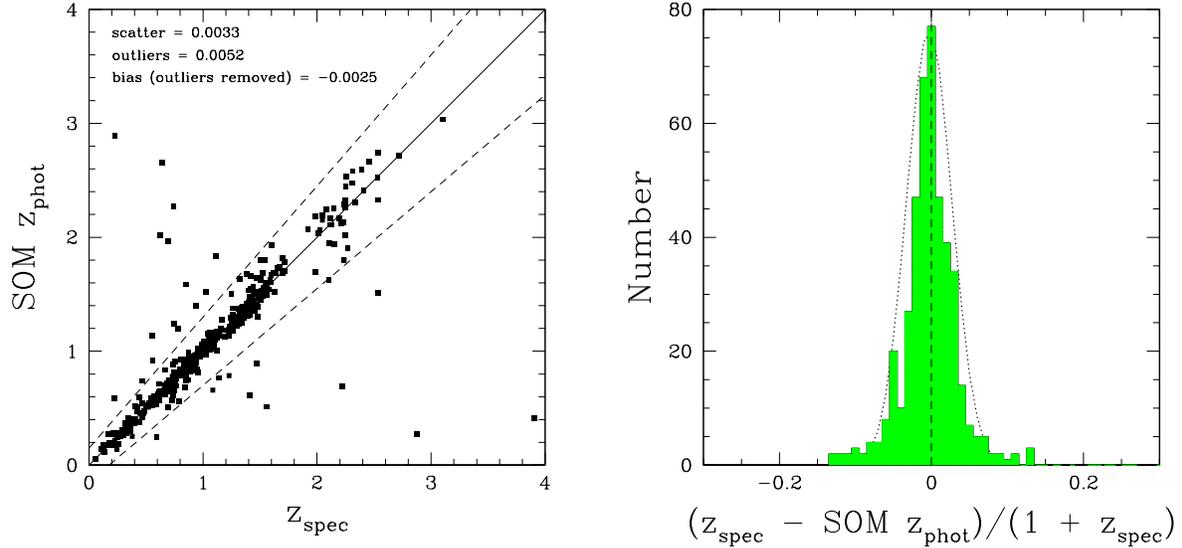

**Figure 2**. The performance of the SOM method on the DR3 redshifts. The plots show all the objects for which high-confidence ($Q = 4$) spectroscopic redshifts were obtained. The true outlier fraction is 0.052, the scatter is 0.033, and after removing outliers the bias is $-0.0025$ over the entire redshift range. In the left panel those objects which are outliers are located outside of the dashed lines. The dotted line in the right panel is a Gaussian centered at zero with $\sigma = 0.033$. Note that the photometry used to place galaxies on the SOM here is generally deeper than the Euclid wide survey, so these results are not fully representative of what Euclid will achieve.

$$\text{mean}\left(\frac{(z_p - z_s)}{1 + z_s}\right)$$

to be $-0.0025$, where the scatter and bias are calculated after removing outliers. If we consider redshifts only in the range $0.2 < z_p < 2.6$, corresponding to the epoch that will be used by *Euclid* for its cosmological weak lensing survey, the bias is $-0.0020$ (after removing outliers). If we include

*good* quality spectroscopic redshifts ($Q = 3$ or $Q = 3.5$) along with the $Q = 4$ redshifts from all three C3R2 data releases, then the scatter is 0.024, the true outlier fraction is 0.056, and the bias is $-0.0018$ (after removing outliers) over the entire redshift range.

In Figure 4, which shows the results for the COSMOS field, we see that there is an increase in the fraction of outliers when all available high-quality spectroscopic redshifts are compared with the SOM photometric redshifts. We dis-cuss the outliers in the DR3 sample in more detail in Section 3.3. Discussion of the other samples from the literature in more detail is beyond the scope of this work; we plan to investigate the causes of the photometric and spectroscopic redshift outliers in the future. Once all the outliers are removed, the bias is calculated to be -0.0020 over the entire redshift range. This improvement relative to the case of only the C3R2 redshifts is likely due to the inclusion of brighter galaxies from the literature samples.

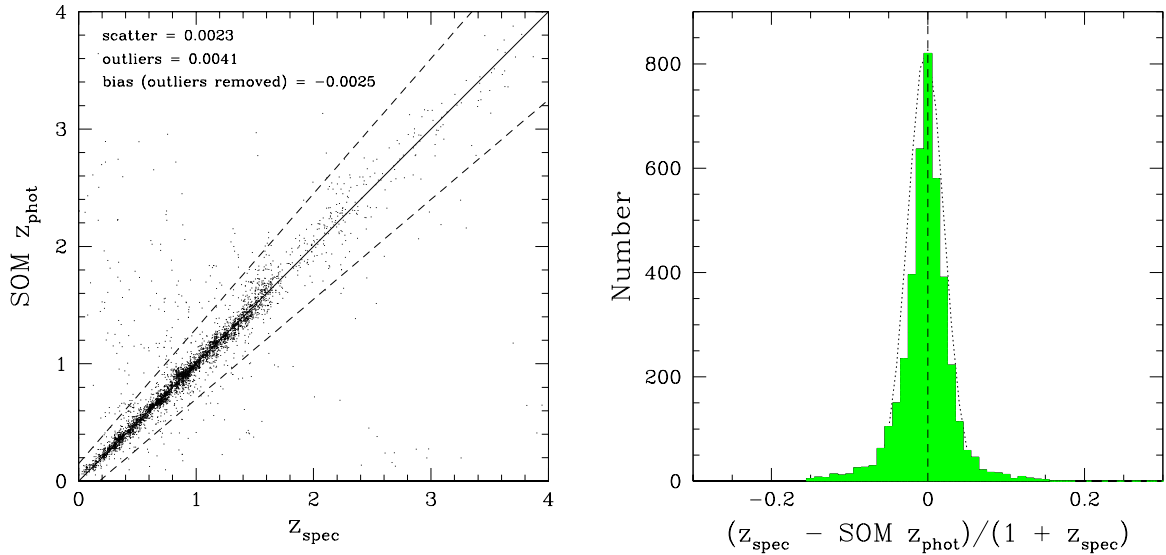

**Figure 3.** Same as in Figure 2, except here we show the performance of the SOM method on the redshifts published by the all three C3R2 data releases. The true outlier fraction for this case is 0.041, the scatter is 0.023, and after removing outliers the bias is −0.0025. The dotted line in the right panel is a Gaussian centered at zero with $\sigma = 0.023$.

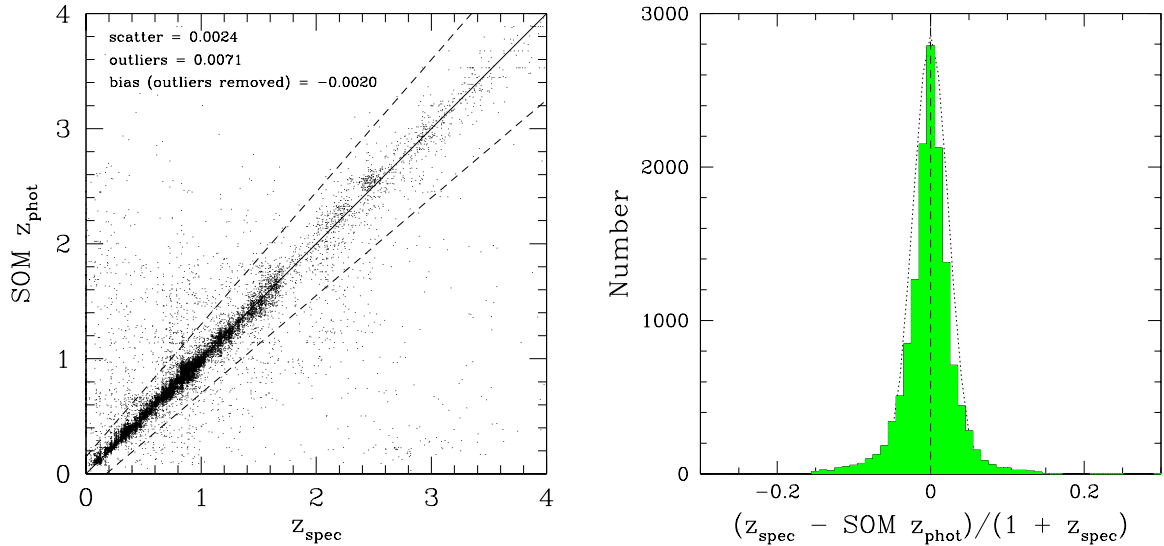

**Figure 4.** Same as in Figure 2, except here we show the performance of the SOM method only in the COSMOS field using all high-quality ($Q = 4$) redshifts from C3R2 combined with all high-quality spectroscopic redshifts available from the literature. For this case, the true outlier fraction is 0.071, the scatter is 0.024, and after removing outliers the bias is −0.0020. The dotted line in the right panel is a Gaussian centered at zero with $\sigma = 0.024$.

In M19 we compared C3R2 spectroscopic redshifts to photometric redshifts in the COSMOS field determined with the template fitting code *Le Phare* (Arnouts et al. 1999; Ilbert et al. 2006) using the same photometry as used here to generate the SOM and SOM-based photometric redshifts. Considering the highest quality ($Q = 4$) redshifts from the three merged C3R2 data releases, these *Le Phare* photometric red-shifts show a scatter $\sigma_{\rm NMAD}$ of 0.029, an outlier fraction of 0.038, and a bias (after removing outliers) of $-0.013$. While the scatter and outlier rate are similar to those from the SOM-based photometric redshifts (Figure 3), the *Le Phare* bias is substantially worse. It may be possible to combine a SOM-determined calibration with the *Le Phare* method in a two step process to produce photometric redshifts for *Euclid*, although it is important to remember that the photometry that will be available for calculating *Euclid* photometric redshifts will not be as good as the existing 30-band photometry in the COSMOS field.

Figure 5 presents the SOMs based on all available high-confidence spectroscopic redshifts in C3R2. The fraction of cells with at least one spectroscopic redshift has increased from 49% in M15 to 84%, and these cells account for 91% of the galaxies. If we restrict consideration to only the $0.2 < z_{\rm p} < 2.6$ range of interest for the *Euclid* cosmological goals, then the fraction of cells with at least one spectroscopic redshift is now 88%, and these cells account for more than 92% of the galaxies.

Figure 6 presents the distributions of the calibrated and uncalibrated cells in the SOM map as a function of magnitude and of redshift. We consider galaxy color cells to be "calibrated" if at least one galaxy in the cell has a high-confidence spectroscopic redshift, while "uncalibrated" cells are those for which no galaxy occupying that cell has a secure redshift measurement. These plots show that most of the uncalibrated cells are faint and located at $z_{\rm p} < 0.6$. The next two sets of plots in Figure 7 further elucidate the nature of these populations. In particular we see that most of the uncalibrated galaxies at faint magnitudes and low photometric redshifts are likely to be passive rather than active, where the rest-frame *UVJ* colors, estimated with the COSMOS2015 (Laigle et al. 2016) data, are used to determine if a galaxy is active or passive. It is worth noting that the photometric redshifts based on *Euclid* and LSST data for faint, passive galaxies at low redshifts are expected to be robust.

### 3.2. *Serendipitous Sources*

In addition to the targeted objects, we also determined $Q = 4$ redshifts for 48 serendipitous sources in DR3, and an additional 19 $Q = 3$ or $Q = 3.5$ redshifts. Some of these serendipitous sources turned out to be in the target catalogs; these are listed in the redshift table with the ID of the matched object and a prefix of "serendip". A few of the serendipitous sources are located very close (e.g., within ∼1 arcsec) to the targeted galaxy. In such situations the catalog photometry is probably affected by blending of the sources within the aperture used on the images to measure brightnesses. This might not be the case for *Euclid* VIS images, where the detection+segmentation will be used to perform template fitting photometry. The extent to which apparently secure redshifts could affect photometric redshift calibration still needs further investigation.

### 3.3. *Assessing the Photometric Redshift Outliers*

As a final step in the analysis, we identified all sources in DR3 with secure redshifts (i.e., quality flag $Q \geq 3$) where the spectroscopic redshift was discrepant from the SOM-based redshift by at least 15% (i.e., $|z_{\rm s} - z_{\rm SOM}|/(1 + z_{\rm s}) > 0.15$). At least two co-authors independently assessed each of the 80 such cases identified.

For three galaxies, this reanalysis lowered the spectroscopic quality flag to $Q < 3$. For the majority of the remaining cases (41/70; 59%), we revised the assigned spectroscopic redshift based on a reconsideration of the data and the photometric redshift. Most of these cases were one-line redshifts initially assigned quality flag $Q = 3$ observed with either MOSFIRE (with its limited spectral range) or LRIS. Typically, a target had a single faint emission line detected, initially assumed to be, for example, H$\alpha$. However, if the line had a different identification, such as [O III], the spectroscopic redshift matched the SOM-based redshift. In all such cases, corroborating lines were searched for, but generally were not found, either due to the limited spectral coverage or strong sky lines. In another two cases, this exercise made us realize that we had misidentified a targeted galaxy with a serendipitous galaxy. In only one case (out of the entire DR3 sample of 635 high-quality redshifts) was a $Q = 4$ redshift corrected by this exercise. In that case, both initial spectroscopic redshift assessments matched, though a re-analysis clearly showed that redshift to be incorrect.

The remaining outliers were approximately evenly split between three categories. For approximately one-third of the outliers, the spectroscopic redshift was discrepant from the SOM-based photometric redshift but matched the *Le Phare* photometric redshift, suggesting an issue with the SOM-based redshift (e.g., potentially a SOM cell with a bimodal photometric redshift distribution). In fact, for 2/11 of these cases, we find that the SOM cell of the galaxies is on a clear "caustic" separating low- and high-redshift galaxies with similar *Euclid* colors, while the others tend to be in cells with intrinsically high redshift dispersion compared with most parts of the color space. It is plausible that the 30-band COSMOS data is capable of resolving these degeneracies, hence the better *Le Phare* photometric redshift estimates. On the other hand, a detailed examination of the *Hubble* ACS imaging of these sources showed that a majority seem to be affected by blending to some degree, which may also contribute to the incorrect SOM-based photometric redshifts.

Another third of the outliers are due to blends. While the input source catalog is based on ground-based, multi-band photometry, in several cases *Hubble* imaging clearly resolved the source into two close galaxies, thereby compromising the photometric redshift. While this was easily assessed in the COSMOS field, such higher resolution imaging was not available for most of the other target fields. Finally, the remaining third of targets were deemed true outliers, where the spectroscopic redshift was deemed solid (e.g., generally $Q = 4$), but did not match either the SOM-based or the *Le Phare* photometric redshift. We expect that several of these cases are also unidentified blends where *Hubble* imaging was not available, highlighting a likely concern for future cosmology projects, both in developing the "gold sample" of calibration photometric redshifts, as well as in assigning pho-tometric redshifts to the large sample of observed objects. It is worth noting that near-Hubble quality imaging will in fact be available for deblending algorithms to use in creating photometric catalogs for *Euclid*.

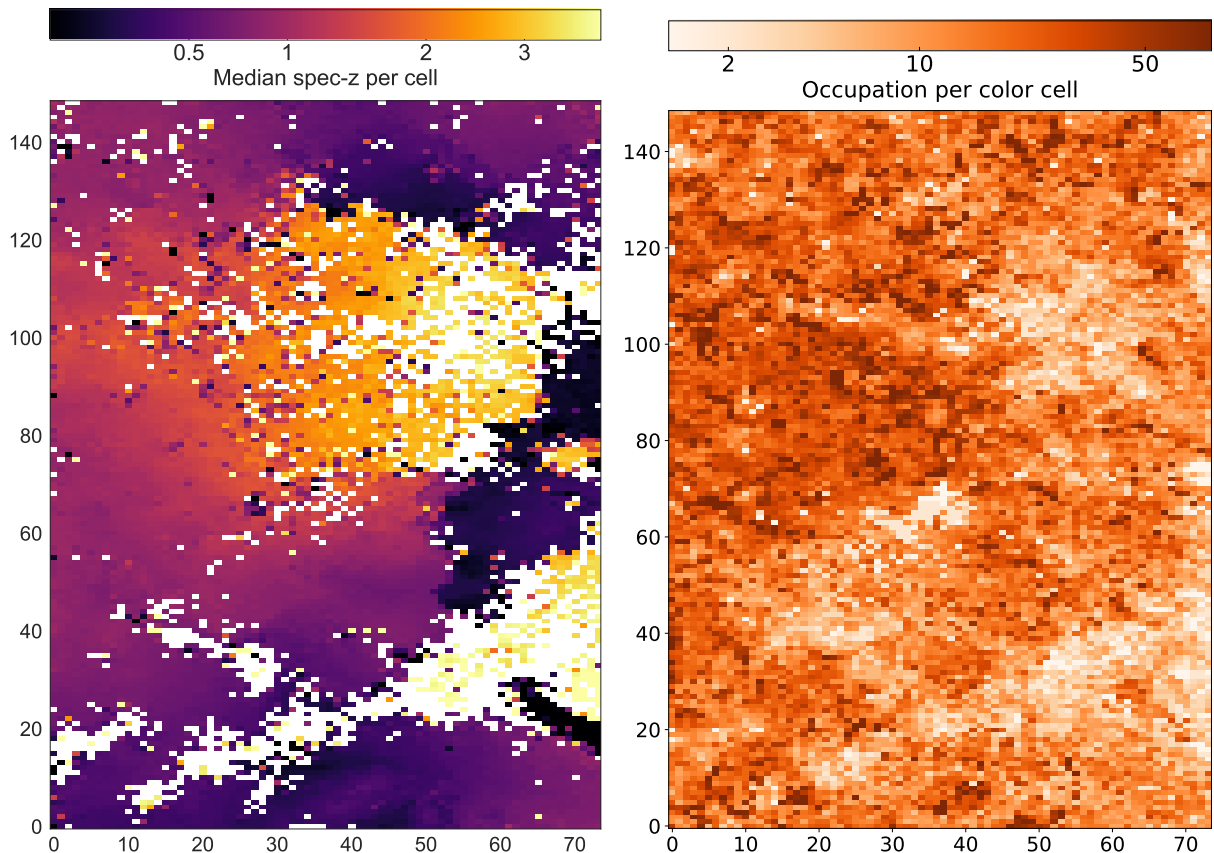

**Figure 5.** Left: The redshift coverage of the SOM color space based on all the $Q \geq 3$ redshifts in C3R2. Right: The current occupation density of the SOM, including the COSMOS and VVDS fields. Approximately 88% of color cells are covered in the range $0.2 < z < 2.6$, and 92% of galaxies live within calibrated cells. The axes are (x,y) indices into the SOM and have no (direct) physical meaning, though a given index corresponds to a particular galaxy SED.

### 3.4. *Spectroscopic Failures*

While we do not expect to spectroscopically confirm all targeted objects, it is useful to examine the possible failure modes that led to not being able to determine a redshift. Neglecting the obvious explanations having to do with poor seeing, clouds, shortened exposures, instrument failure (one LRIS night, primarily targeting Lyman-break galaxies, was conducted without the blue side), there were 32 masks out of the 50 observed which were observed in at least good conditions. These 32 good masks had 1346 targeted objects, of which no secure redshift was obtained for 835 sources. Because the COSMOS catalog has the best photometry and so should have the best predicted exposure times, we restrict our investigation of the failures to only the 597 objects observed in the COSMOS field. The distribution of the good COSMOS masks by instrument is 10/5/1 for MOS-FIRE/DEIMOS/LRIS, respectively. There were 70 objects for which a low-confidence ($Q = 1$ or 2) redshift was determined, 162 with a failure code of $-91$ (too faint), 39 with a failure code of $-92$ (reasonable detection but no redshift), and 25 have a code of $-93$ or $-94$ meaning there were issues with the data or reductions. The success rates by instrument are fairly similar for DEIMOS (43%) and MOSFIRE (37%) for the good COSMOS masks; there was only one mask done by LRIS in good conditions in the COSMOS field.

The success rate for the good LRIS observations in all three fields was lower compared to that obtained with DEIMOS and MOSFIRE. High confidence redshifts were obtained of only 61 of the 413 targeted objects in the 14 good LRIS masks; no secure redshifts were obtained for 347 objects, and the reductions were problematic for five objects. The high confidence redshift success rate for the good LRIS masks was 15%. The vast majority of the targeted objects in the LRIS masks are $1.6 < z_p < 3$ Lyman-break galaxies which often do not have any emission lines. In these cases, redshifts usually need to be determined from rest frame UV absorption features which are much more difficult to detect. This analysis suggests that for the $z \sim 2$ Lyman-break galaxies we should revise the way that we predict exposure times. Possible solutions to the low success rate with the LRIS observations of such galaxies would be to obtain deeper exposures, or to try observing such targets with MOSFIRE which offers the possibility of detecting emission lines in the rest frame optical.

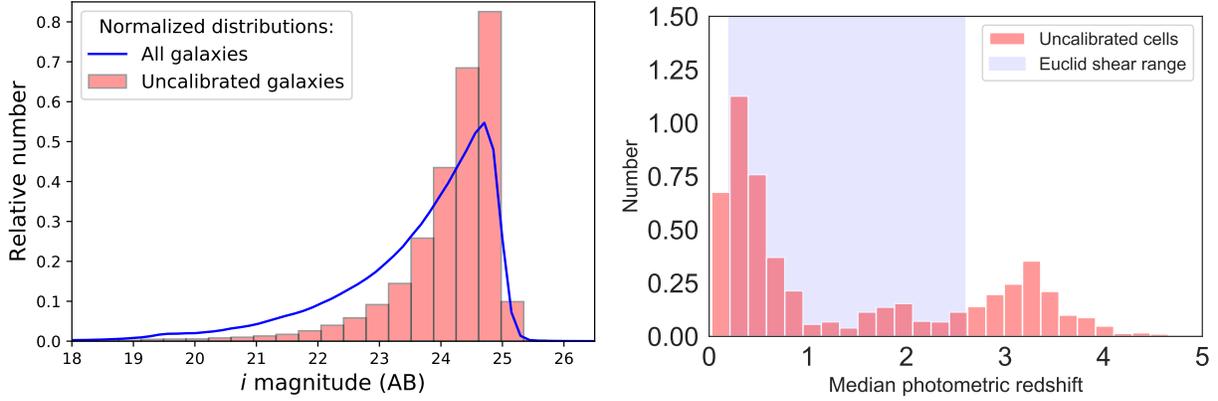

**Figure 6.** Left: The histogram of the *i*-band median magnitudes of cells in the SOM, with all cells shown by the blue curve and the uncalibrated cells (those which do not yet have spectroscopic redshifts) in red. Right: The photometric redshift histogram of the uncalibrated cells in the SOM. The blue shading delineates the redshift range of the objects that will be used for *Euclid* shear measurements.

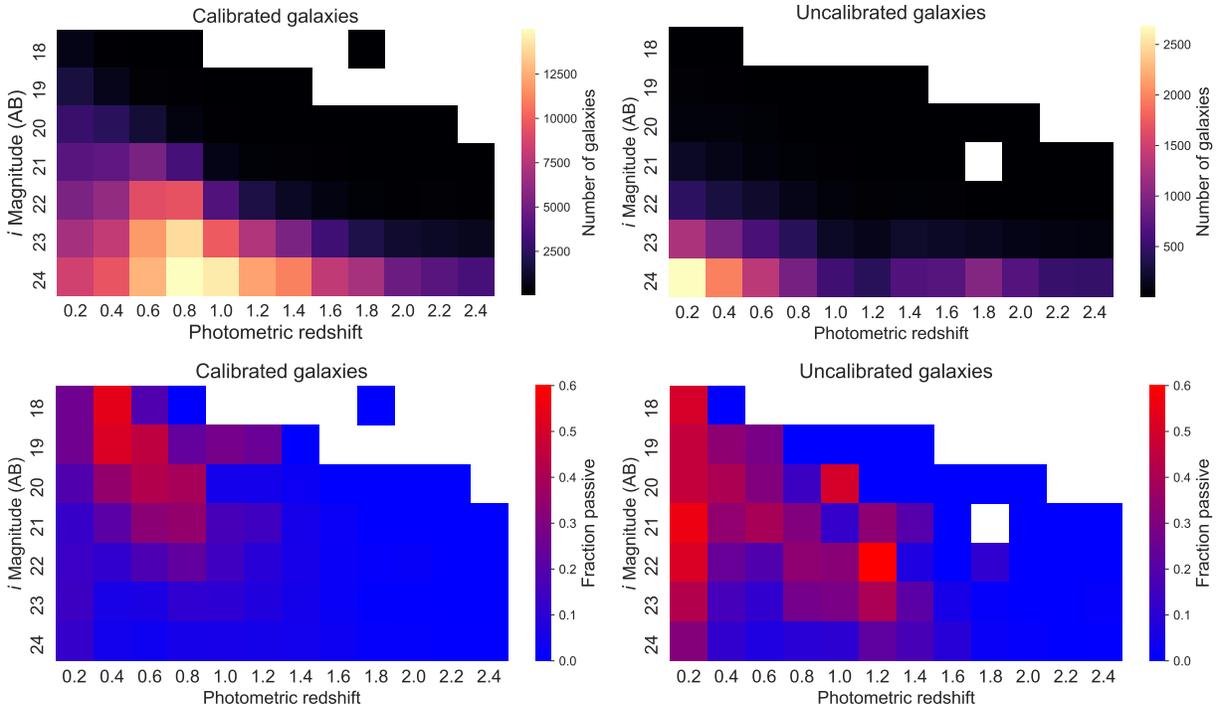

**Figure 7.** Top left: The distribution of "calibrated" galaxies (i.e., those in SOM cells with at least one spectroscopic redshift) in the plane of redshift and *i*-band magnitude. Top right: The same plot as in the top right panel, but for currently "uncalibrated" galaxies. Bottom left: The relative fractions of passive galaxies in the calibrated set as a function of magnitude and redshift, as estimated from rest-frame *UVJ* colors. Bottom right: The same plot but for the uncalibrated galaxies. It is clear from these figures that the currently uncalibrated galaxies are fainter and tend to be more quiescent than the calibrated galaxies. In all panels the analysis uses galaxies in the COSMOS2015 sample, and white areas represent no galaxies in a bin.

## 4. SUMMARY

We have presented 676 new high-confidence spectroscopic redshifts obtained by the C3R2 survey. Taken together with the previous two data releases, we have now collected and published 5130 high-confidence spectroscopic redshifts (and associated publicly released spectra) covering a wide range in galaxy type and redshift.[2]

While good progress has been made in covering the *Euclid* color space with spectroscopic redshifts, and the prospects for using these redshifts to calibrate photo-z calculations for *Euclid* are promising, challenges remain. More work remains to be done to validate published samples of spectroscopic redshifts before they can also be used for this purpose. Also, the relatively low success rate of the observations (particularly with LRIS) of galaxies without emission lines emphasizes the technical challenges faced by redshift surveys such as C3R2 that are attempting to calibrate photometric redshifts with great precision down to the faint magnitudes of interest to upcoming space missions such as *Euclid*.

A significant portion of the research and manuscript preparation took place during the COVID-19 global pandemic. The authors would like to thank all those who risked their lives as essential workers in order for us to safely continue our work from home. The authors also thank the referee whose comments helped improve the clarity and presentation of our results.

The research was carried out in part at the Jet Propulsion Laboratory, California Institute of Technology, under a contract with the National Aeronautics and Space Administration. This work was enabled by a NASA Keck grant. FJC acknowledges support from grant ESP2017-89838 ad H2020 European Commission grant 776247.

The data presented herein were obtained at the W. M. Keck Observatory, which is operated as a scientific partnership among the California Institute of Technology, the University of California and the National Aeronautics and Space Administration. The Observatory was made possible by the generous financial support of the W. M. Keck Foundation. The authors wish to recognize and acknowledge the very significant cultural role and reverence that the summit of Maunakea has always had within the indigenous Hawaiian community. We are most fortunate to have the opportunity to conduct observations from this mountain.

The Euclid Consortium acknowledges the European Space Agency and a number of agencies and institutes that have supported the development of *Euclid*, in particular the Academy of Finland, the Agenzia Spaziale Italiana, the Belgian Science Policy, the Canadian Euclid Consortium, the Centre National d'Etudes Spatiales, the Deutsches Zentrum für Luft- und Raumfahrt, the Danish Space Research Institute, the Fundação para a Ciência e a Tecnologia, the Ministerio de Economia y Competitividad, the National Aeronautics and Space Administration, the Netherlandse Onderzoekschool Voor Astronomie, the Norwegian Space Agency, the Romanian Space Agency, the State Secretariat for Education, Research and Innovation (SERI) at the Swiss Space Office (SSO), and the United Kingdom Space Agency. A complete and detailed list is available on the *Euclid* web site (http://www.euclid-ec.org).

## APPENDIX

Table 4 is published in its entirety in machine-readable format. A portion is shown here for guidance regarding its form and content. More detailed information is available in the header of the machine-readable table.

**Table 4**. C3R2 DR3 Spectroscopic Redshifts

| Object ID | R.A. (J2000) | Dec (J2000) | Mask | Slit | $i$ mag | $z$ | Q | Instr. | Filename |
|---|---|---|---|---|---|---|---|---|---|
| VVDS-103460 | 02 27 32.10 | -04 47 41.6 | D094-VVDS32 | 15 | 24.09 | 0.5047 | 4.0 | DEIMOS | spec1d.vvds32.015.VVDS-103460.fits |
| VVDS-98275 | 02 27 06.64 | -04 48 14.5 | D094-VVDS32 | 16 | 24.38 | 1.0881 | 4.0 | DEIMOS | spec1d.vvds32.016.VVDS-98275.fits |
| VVDS-107542 | 02 27 35.45 | -04 47 14.3 | D094-VVDS32 | 25 | 24.34 | 0.8819 | 4.0 | DEIMOS | spec1d.vvds32.025.VVDS-107542.fits |
| VVDS-121353 | 02 26 59.16 | -04 45 39.2 | D094-VVDS32 | 38 | 22.24 | 0.6153 | 4.0 | DEIMOS | spec1d.vvds32.038.VVDS-121353.fits |
| VVDS-89949 | 02 27 29.90 | -04 49 09.7 | D094-VVDS32 | 61 | 24.02 | 0.8398 | 4.0 | DEIMOS | spec1d.vvds32.061.VVDS-89949.fits |
| VVDS-125603 | 02 27 08.15 | -04 45 04.7 | D094-VVDS32 | 62 | 24.86 | 0.8691 | 4.0 | DEIMOS | spec1d.vvds32.062.VVDS-125603.fits |
| VVDS-120554 | 02 26 59.91 | -04 45 43.0 | D094-VVDS32 | 63 | 24.78 | 0.9038 | 4.0 | DEIMOS | spec1d.vvds32.063.VVDS-120554.fits |
| VVDS-118835 | 02 27 14.55 | -04 45 53.9 | D094-VVDS32 | 64 | 24.59 | 0.9076 | 4.0 | DEIMOS | spec1d.vvds32.064.VVDS-118835.fits |
| VVDS-112287 | 02 27 46.85 | -04 46 39.6 | D094-VVDS32 | 67 | 24.86 | 1.2580 | 4.0 | DEIMOS | spec1d.vvds32.067.VVDS-112287.fits |
| VVDS-97940 | 02 27 01.46 | -04 48 17.7 | D094-VVDS32 | 69 | 24.64 | 0.9837 | 4.0 | DEIMOS | spec1d.vvds32.069.VVDS-97940.fits |
| VVDS-110496 | 02 26 53.92 | -04 46 53.3 | D094-VVDS32 | 71 | 24.82 | 1.1917 | 4.0 | DEIMOS | spec1d.vvds32.071.VVDS-110496.fits |
| VVDS-111705 | 02 27 52.70 | -04 46 44.1 | D094-VVDS32 | 72 | 24.46 | 0.4468 | 4.0 | DEIMOS | spec1d.vvds32.072.VVDS-111705.fits |